\documentclass[12pt]{iopart}
\usepackage{rotating,epsf }

\newcommand{\nn}{\nonumber}

\newcommand{\alink}{\mbox{
\begin{picture}(2.5,.2)
\linethickness{1mm}
\multiput(0,0.1)(2,0){2}{\circle*{0.1}}
\multiput(1.01,0.1)(0,0){1}{\circle{0.2}}
\put(0.8,0.){\link}
\put(-0.2,0.){\linka}
\put(1.1,-.2){\scriptsize \( x \) }
\put(0,-.2){\scriptsize \( y \) }
\put(2.0,-.2){\scriptsize \( y \) }
\end{picture}}}
\newcommand{\alinkfat}{\mbox{
\begin{picture}(3.0,.2)
\linethickness{1mm}
\multiput(0,0.1)(2.4,0){2}{\circle*{0.1}}
\multiput(1.21,0.1)(0,0){1}{\circle{0.2}}
\put(1.2,-1.){\staple}
\put(-0.2,-1.){\staplea}
\put(1.1,-.3){\scriptsize \( x \) }
\put(-.4,-.2){\scriptsize \( y \) }
\put(2.6,-.2){\scriptsize \( y \) }
\end{picture}}}
\newcommand{\blinkba}{\mbox{
\begin{picture}(2.5,2.5)
\thicklines
\multiput(1,-1)(0,-1){2}{\circle*{0.1}}
\multiput(1,0)(0,0){1}{\circle{0.2}}
\multiput(2,0.0)(0,1){3}{\circle*{0.1}}
\multiput(0,-2)(0,0){1}{\circle*{0.1}}
\put(1,-2.0){\vector(-1,0){1}}
\put(1,-1.0){\vector(0,-1){1}}
\put(1,0.0){\vector(0,-1){1}}
\put(1,0.0){\vector(1,0){1}}
\put(2,0.0){\vector(0,1){1}}
\put(2,1.0){\vector(0,1){1}}
\put(0.5,-0.1){\scriptsize \( x \) }
\put(1.5,2.0){\scriptsize \( y \) }
\put(-0.4,-2.1){\scriptsize \( y \) }
\end{picture}}}
\newcommand{\blinkbb}{\mbox{
\begin{picture}(2.5,2.5)
\thicklines
\multiput(0,0.0)(0,-1){3}{\circle*{0.1}}
\multiput(1,1.0)(0,1){2}{\circle*{0.1}}
\multiput(1,0.0)(0,){1}{\circle{0.2}}
\multiput(2,2)(0,1){1}{\circle*{0.1}}
\put(0,-1.0){\vector(0,-1){1}}
\put(0,0){\vector(0,-1){1}}
\put(1,0){\vector(-1,0){1}}
\put(1,0){\vector(0,1){1}}
\put(1,1){\vector(0,1){1}}
\put(1,2){\vector(1,0){1}}
\put(1.25,-0.1){\scriptsize \( x \) }
\put(2.2,2.0){\scriptsize \( y \) }
\put(-0.4,-2.1){\scriptsize \( y \) }
\end{picture}}}
\newcommand{\blinkbc}{\mbox{
\begin{picture}(2.5,2.5)
\thicklines
\multiput(1,1)(0,1){2}{\circle*{0.1}}
\multiput(1,0)(0,0){1}{\circle{0.2}}
\multiput(2,0)(0,-1){3}{\circle*{0.1}}
\multiput(0,2)(0,1){1}{\circle*{0.1}}
\put(1,2){\vector(-1,0){1}}
\put(1,1){\vector(0,1){1}}
\put(1,0){\vector(0,1){1}}
\put(1,0){\vector(1,0){1}}
\put(2,0){\vector(0,-1){1}}
\put(2,-1){\vector(0,-1){1}}
\put(0.5,-0.1){\scriptsize \( x \) }
\put(1.5,-2.1){\scriptsize \( y \) }
\put(-0.5,2.0){\scriptsize \( y \) }
\end{picture}}}
\newcommand{\blinkbd}{\mbox{
\begin{picture}(2.5,2.5)
\thicklines
\multiput(0,0)(0,1){3}{\circle*{0.1}}
\multiput(1,-1)(0,-1){2}{\circle*{0.1}}
\multiput(1,0)(0,-1){1}{\circle{0.2}}
\multiput(2,-2)(0,1){1}{\circle*{0.1}}
\put(0,1){\vector(0,1){1}}
\put(0,0){\vector(0,1){1}}
\put(1,0){\vector(-1,0){1}}
\put(1,0){\vector(0,-1){1}}
\put(1,-1){\vector(0,-1){1}}
\put(1,-2){\vector(1,0){1}}
\put(1.25,-0.1){\scriptsize \( x \) }
\put(2.1,-2.1){\scriptsize \( y \) }
\put(-0.5,2.0){\scriptsize \( y \) }
\end{picture}}}
\newcommand{\staple}{\mbox{
\begin{picture}(1.2, 2.2)
\thicklines
\put(0,1.2){\vector(0,1){1}}
\put(0,1){\vector(0,-1){1}}
\put(0,2.2){\vector(1,0){1}}
\put(0,0.){\vector(1,0){1}}
\put(1,0.){\vector(0,1){1}}
\put(1,2.2){\vector(0,-1){1}}
\end{picture}}}
\newcommand{\staplea}{\mbox{
\begin{picture}(1.2, 2.2)
\thicklines
\put(1,1.2){\vector(0,1){1}}
\put(0,2.2){\vector(0,-1){1}}
\put(1,2.2){\vector(-1,0){1}}
\put(1,1){\vector(0,-1){1}}
\put(1,0){\vector(-1,0){1}}
\put(0,0){\vector(0,1){1}}
\end{picture}}}
\newcommand{\link}{\mbox{
\begin{picture}(1.1, .1)
\thicklines
\put(0,0.1){\vector(1,0){1}}
\end{picture}}}
\newcommand{\linka}{\mbox{
\begin{picture}(1.1, .1)
\thicklines
\put(1,0.1){\vector(-1,0){1}}
\end{picture}}}

\newsavebox{\Staple}
\savebox{\Staple}{\begin{picture}(0,0)
\thicklines
\put(0.0,0.1){\vector(0,1){0.9}}
\put(0.0,1.0){\vector(1,0){0.9}}
\put(0.9,1.0){\vector(0,-1){0.9}}
\end{picture}}

\newsavebox{\InvertedStaple}
\savebox{\InvertedStaple}{\begin{picture}(0,0)
\thicklines
\put(0.0,-0.1){\vector(0,-1){0.9}}
\put(0.0,-1.0){\vector(1,0){0.9}}
\put(0.9,-1.0){\vector(0,+1){0.9}}
\end{picture}}

\newsavebox{\FiveStaple}
\savebox{\FiveStaple}{\begin{picture}(0,0)
\thicklines
\put(0.0,0.1){\vector(0,1){0.9}}
\put(0.0,1.0){\vector(1,1){0.5}}
\put(0.5,1.5){\vector(1,0){0.9}}
\put(1.4,1.5){\vector(-1,-1){0.5}}
\put(0.9,1.0){\vector(0,-1){0.9}}
\end{picture}}

\newsavebox{\SevenStaple}
\savebox{\SevenStaple}{\begin{picture}(0,0)
\thicklines
\put(0.0,0.1){\vector(0,1){0.9}}
\put(0.0,1.0){\vector(1,1){0.5}}
\put(0.5,1.5){\vector(1,2){0.3}}
\put(0.8,2.1){\vector(1,0){0.9}}
\put(1.7,2.1){\vector(-1,-2){0.3}}
\put(1.4,1.5){\vector(-1,-1){0.5}}
\put(0.9,1.0){\vector(0,-1){0.9}}
\end{picture}}

\newsavebox{\LepageStaple}
\savebox{\LepageStaple}{\begin{picture}(0,0)
\thicklines
\put(0.0,0.1){\vector(0,1){0.9}}
\put(0.0,1.0){\vector(0,1){0.9}}
\put(0.0,1.9){\vector(1,0){0.9}}
\put(0.9,1.9){\vector(0,-1){0.9}}
\put(0.9,1.0){\vector(0,-1){0.9}}
\end{picture}}

\newsavebox{\Link}
\savebox{\Link}{\begin{picture}(0,0)
\thicklines
\put(0.0,0.0){\vector(1,0){0.9}}
\end{picture}}

\newsavebox{\Naik}
\savebox{\Naik}{\begin{picture}(0,0)
\thicklines
\put(0.0,0.0){\vector(1,0){0.9}}
\put(1.0,0.0){\vector(1,0){0.9}}
\put(2.0,0.0){\vector(1,0){0.9}}
\end{picture}}

\begin{document}

\title{Lattice results with three quark flavours}
\author{
Steven Gottlieb}
\address{Department of Physics SW117, Indiana University, Bloomington IN 47405,
USA}

\begin{abstract}
There have been exciting advances recently in finite temperature
calculations with three quark flavours and with non-zero chemical potential.
The role of improved actions is explained and recent results from the 
Bielefeld group and MILC collaboration with improved Kogut-Susskind
quarks are presented.  Three new approaches to finite chemical potential
are discussed.

\end{abstract}
\pacs{11.15Ha, 05.70.Ce, 12.38.Gc, 12.38.Mh}

\section{Introduction}
I have been asked by the organizers to review the status of lattice QCD
calculations at non-zero temperature with three quark flavours.  
The exciting news here is that due to improvements in algorithms, we
are now able to carry out calculations with light up and down quarks
considerably lighter than the strange quark.  Before
presenting results, I give a brief introduction to lattice calculations,
and then I will outline how algorithms have been improved.  After
presenting numerical results, I will discuss briefly the recent significant
advances in dealing with finite chemical potential.  (I ran out of time
during my talk and did not discuss this at the conference.)

\section{Introduction to Lattice Calculations}

To carry out a lattice simulation we must select certain parameters, namely
the lattice spacing ($a$) or gauge coupling ($\beta$),
a fixed grid size ($N_s^3 \times N_t$) where $N_s$ and $N_t$ are the
space and time dimensions of the grid, respectively, and
quark masses ($m_{u,d}$, $m_s$).  There are also certain parameters related
to the algorithm.  The parameters detailed have physical meaning, and
each choice can lead to a systematic error that must be controlled.

To deal with the non-zero lattice spacing, we
must take the continuum limit.  Similarly, we 
must take the infinite volume limit to deal with the finite fixed grid.
Finally, we must extrapolate to light quark mass for the up and down quarks. 
This is a practical issue as it is too expensive to do the calculations
with the physical up and down masses.  On the other hand, we
can work at the physical $s$ quark mass.
For finite temperature calculations, the temperature is
given by $T=1/(N_t a)$, so we use grids with
$N_t < N_s$.  Typically, $N_s$ is two or three times $N_t$ to provide a
reasonably large physical volume.

Also of physical relevance (especially for the RHIC program) is
nonzero chemical potential $\mu$.  Until recently, it has been nearly
impossible to carry out calculations with $\mu\ne0$.

\section{How have algorithms been improved?}

The first numerical lattice calculations used the Wilson (plaquette)
gauge action and either Wilson or Kogut-Susskind (KS or staggered) 
quark actions.
The Wilson quark action is designed to solve the fermion ``doubling'' problem.
So is the KS approach; however, it maintains enough degrees of
freedom for four quark fields.  There remains a $U(1)$ chiral symmetry
and a single Goldstone pion.  The rest of the flavour symmetry is broken for
non-zero lattice spacing.
Thus, the other pion states are heavier than the Goldstone state because of
flavour (or what is now frequently called taste) symmetry breaking.

Starting in the mid-1980s, the Symanzik improvement program 
\cite{Symanzik:1983dc}
began to be
applied to calculations to improve the scaling properties of the
theory.  
For Wilson type quarks the improvement called clover quarks was introduced
by Sheikholeslami and Wohlert \cite{Sheikholeslami:1985ij}.
Naik introduced a new three link term for KS quarks in 1989 \cite{Naik:1989}.
An improved gauge action was applied in 1995 \cite{Alford:1995hw}.
Soon thereafter calculations were done with the Naik term and an improved
gauge action \cite{Bernard:1996em}
and the importance of ``fattening'' the action to reduce taste symmetry
breaking was tested and understood [6--10]. 
About the same time, the P4 action was introduced because of it's improved
rotational symmetry \cite{Engels:1996ag,Heller:1999xz}

All of these improvement schemes are 
analagous to higher order methods in numerical analysis, but with the
twist of applying to QFT.  They exact a computational toll from having a
more complicated action, but they more than make up for that by reducing
systematic errors and allowing a larger lattice spacing to be used.  Since
computational requirement goes like a large power (7--9) of the inverse lattice
spacing, it really reduces the cost to use a larger lattice spacing.

More recently, 
domain wall fermions and the overlap method
were developed to better control chiral symmetry. 
They are not yet extensively used
for thermodynamics, but see reference \cite{Chen:2000zu}.

We shall concentrate on the improvement program for staggered quarks.
The Bielefeld group is the center of activity for the
P4 action that maintains rotational
invariance of the free quark propagator to order $p^4$,
 
\unitlength0.35cm
\begin{eqnarray}
\lefteqn{ S_F (m_{f,L}) ~=~ c_1^F S_{1-link,fat} (\omega) +
c_3^F S_{3-link}+  m_{f,L}\sum_x
\bar{\chi}_x^f \chi_x^f \nn ~~~~~~~~~~~~~~~~~~~~~~~}\\
&\equiv& \sum_x \bar{\chi}_x^f~\sum_\mu ~ \eta_\mu(x) ~ \Bigg(
{3\over 8}~\Bigg[ \alink~ +~ \omega~~\sum_{\nu \ne \mu}~~ \alinkfat\Bigg]
\nn \\[4mm]
& & + {1\over96}~\sum_{\nu\ne \mu} ~\Bigg[ \blinkbd + \blinkbc ~+
 \blinkba + \blinkbb \Bigg] \Bigg) \chi_y^f \nn \\[13mm]
& & + m_{f,L}  \sum_x~\bar{\chi}_x^f \chi_x^f  \quad .
\end{eqnarray}

\begin{figure}[htb]
\vspace{-0.1in}
\setlength{\unitlength}{0.75in}
\begin{center}\begin{picture}(6.0,2.0)
\put(0.0,0){\usebox{\Link}\makebox(0,0)}
\put(1.0,0){\usebox{\Staple}\makebox(0,0)}
\put(2.0,0){\usebox{\FiveStaple}\makebox(0,0)}
\put(3.0,0){\usebox{\SevenStaple}\makebox(0,0)}
\put(4.5,0){\usebox{\LepageStaple}\makebox(0,0)}
\end{picture}\end{center}
\caption{
   \label{PATHSET}
Terms used to suppress flavour symmetry breaking.  The final
   five link path was introduced by Lepage to correct the
   small momentum form factor.
}
\end{figure}
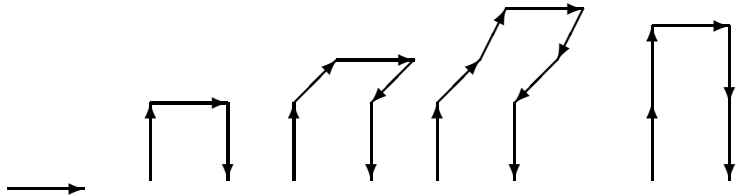

The MILC collaboration uses the ``Asqtad'' action
\cite{Lepage:1998vj,Orginos:1999cr,Bernard:1999xx}.
Their gauge action includes plaquette,
$1\times 2$ and bent 6-link terms.  Their quark action
includes the links shown below and the 3-link Naik term.
The three, five and seven link terms are known as fat links and are
necessary to reduce taste symmetry breaking.
The full set of fattened links is pictured in \fref{PATHSET}..
The weights of each term are not shown.  Each diagram represents a term
in the quark action of the form $\bar\chi(x) V(x,x+\hat\mu) \chi(x+\hat\mu)$,
where $V$ is the product of links along the path.
With coefficients of the various terms set using tadpole improvement,
the errors of this action are of order $a^2 \alpha$ and $a^4$.

\begin{figure}
\begin{center}
\begin{tabular}{c c}
\includegraphics[width=2.35in]{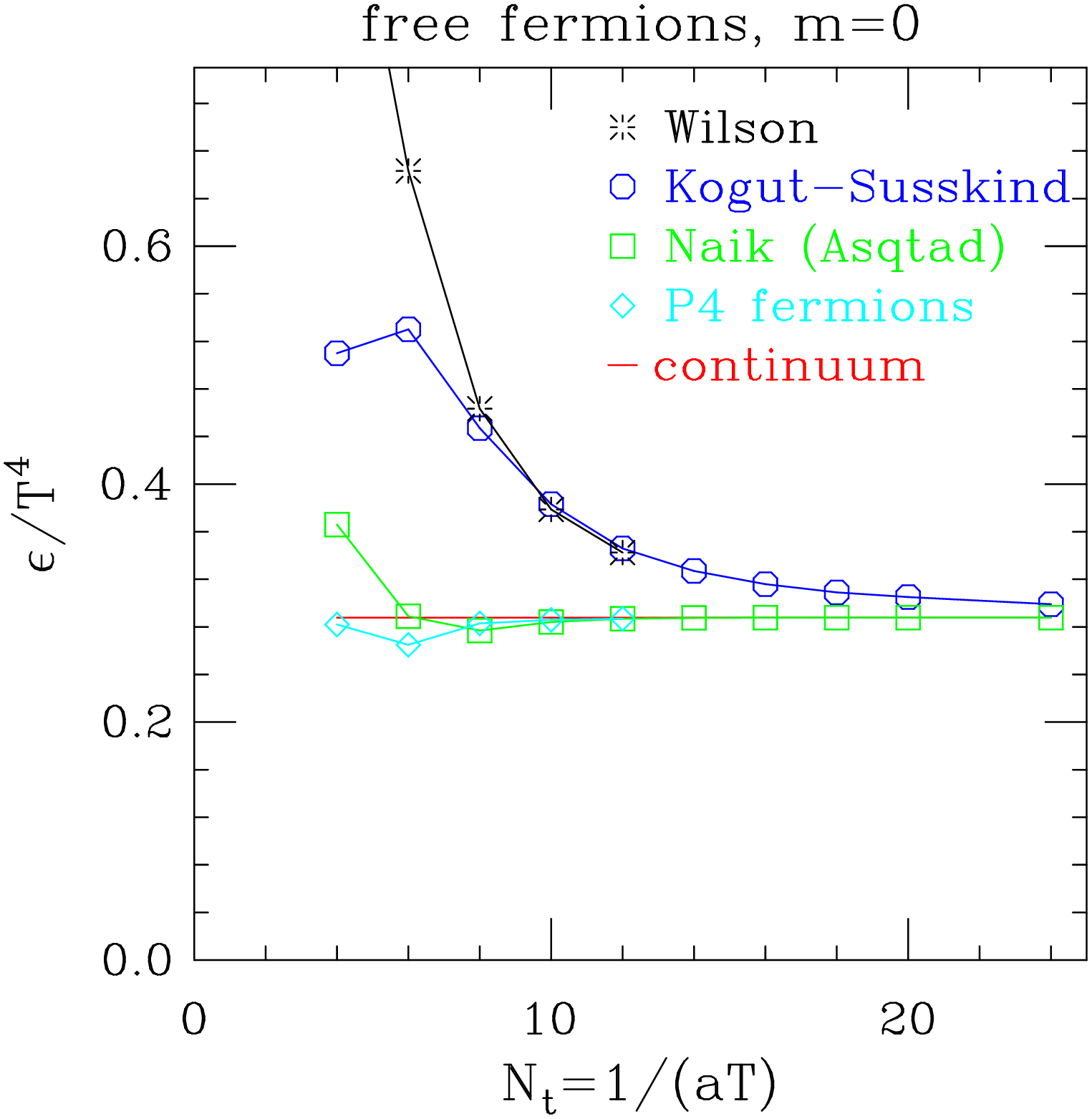}
&
\includegraphics[width=2.35in]{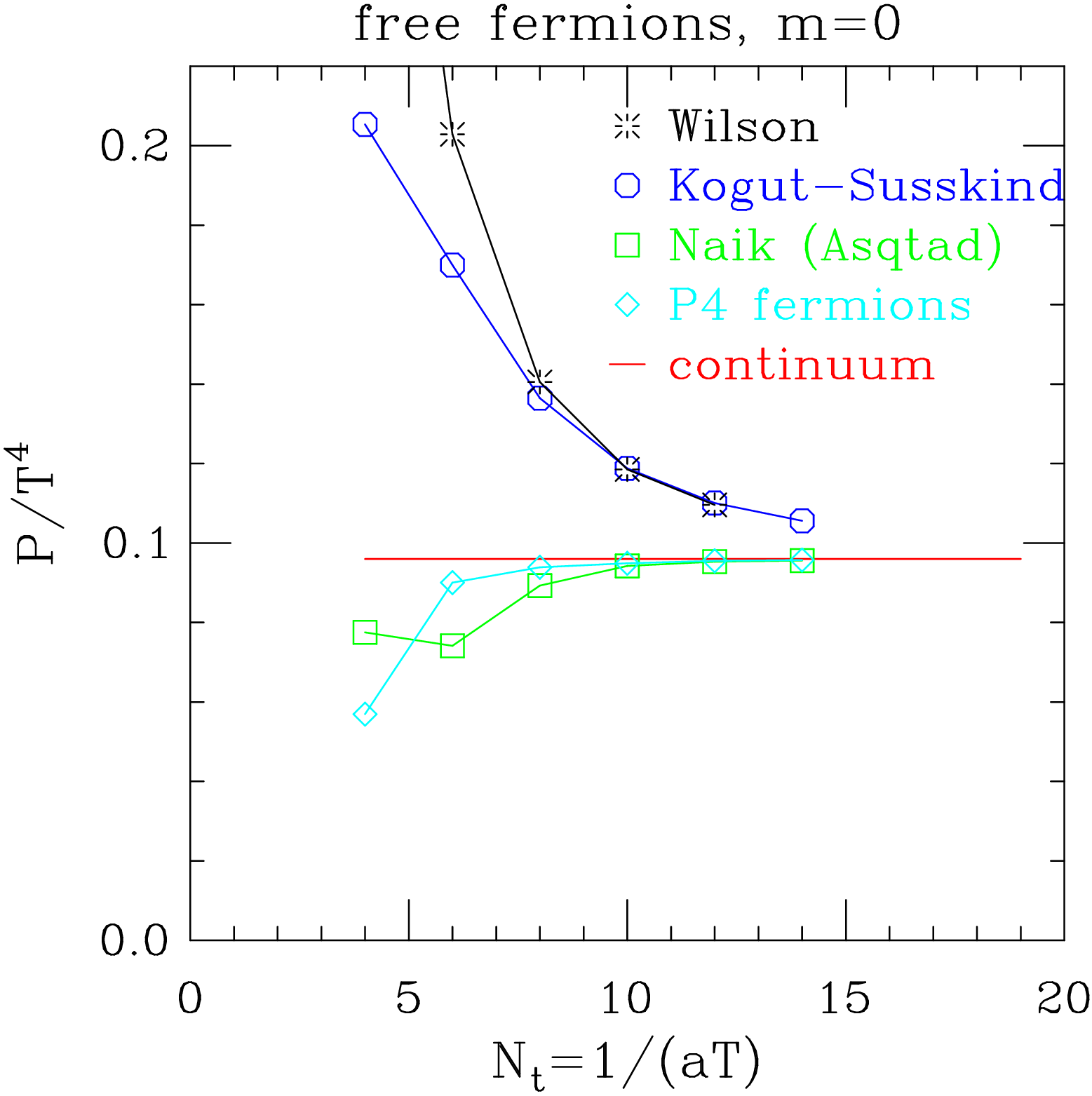}
\end{tabular}
\end{center}
\caption{The energy density (left) and pressure (right) of free massless
fermions as function of temporal lattice size $N_t$. 
}
\label{free_e_p}
\end{figure}

It is easy to see why 
improved actions should have better high temperature behavior than
the Wilson or KS actions.  In \fref{free_e_p}, we can examine
the energy and pressure for free massless quarks on a lattice with $N_t$
time slices \cite{Bernard:2001fi}.  The continuum limit corresponds to $N_t\rightarrow\infty$.
One immediately sees that both Naik and P4 actions much more rapidly approach
the continuum limit than either Wilson or KS quarks.  At $N_t=4$,
Naik is better for pressure and P4 is better for energy density.  For $N_t=6$,
the situation is reversed.  For $N_t\ge8$, both improved actions are quite
close to the continuum limit, with P4 a little closer.

\section{Some recent physics results}

We will now
concentrate on results with 3 degenerate or 2+1 flavours of quark.  Some of
the issues to be addressed are: a) what is the phase diagram of QCD?, 
b) what is the transition temperature?, c) where does the the physical
point in the phase diagram lie?, d) what is the equation of state for QCD?

\begin{figure}[tb]
\epsfxsize=4.0in
\epsfysize=4.5in
\vspace{-1.5in}
\begin{center}
\epsfbox{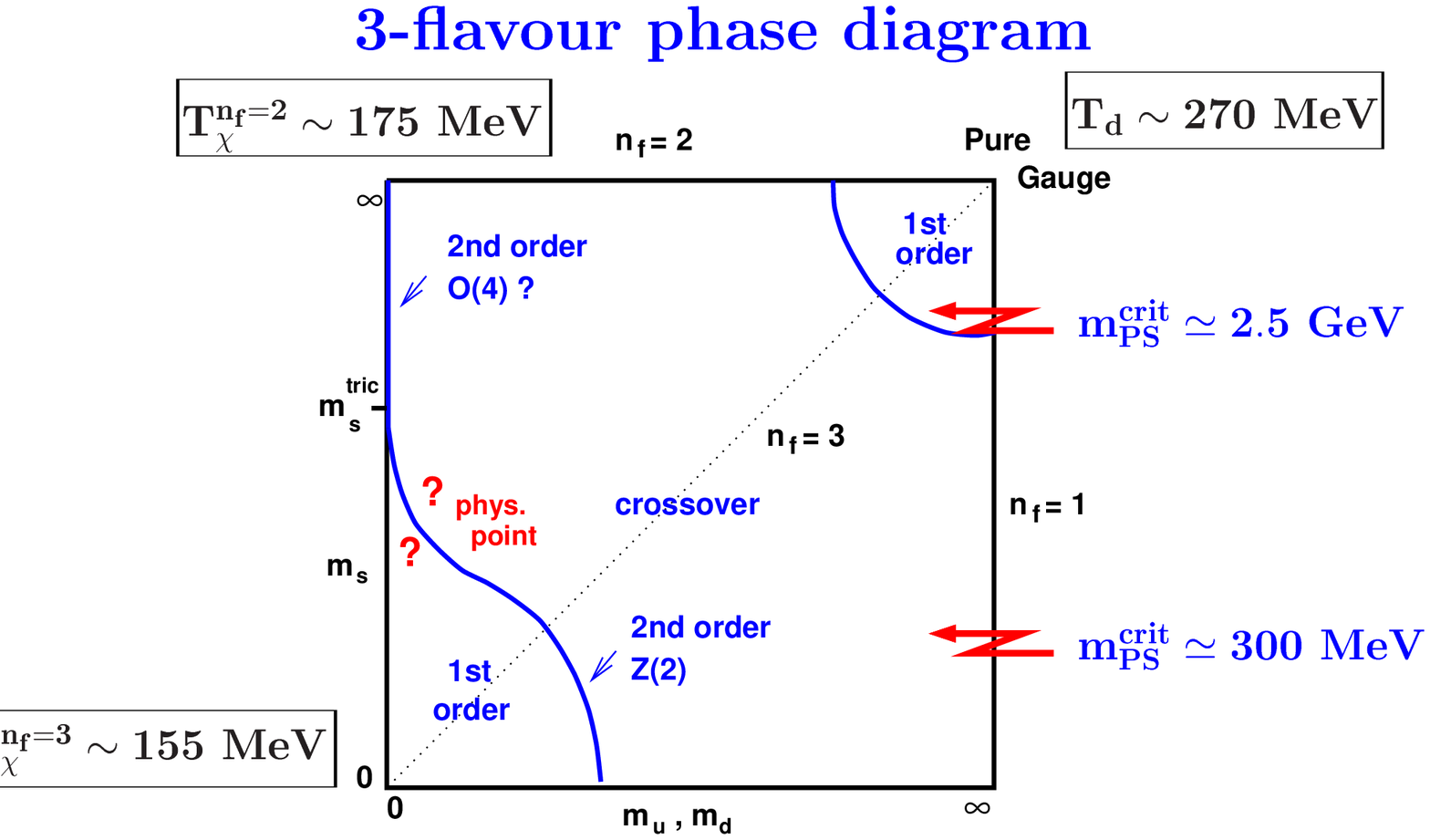}
\end{center}
\vspace{-1.00in}
\caption{From F. Karsch \cite{Karsch:2001vs}.}
\label{phase_diagram}
\end{figure}

To find the phase diagram, we must find the transition temperature for
a variety of quark masses, so it would be logical to address (b) before (a);
however, presenting the phase diagram first helps to orient us, so in 
\fref{phase_diagram}, 
I present a nice summary plot prepared by Karsch of the QCD 
phase diagram in the $m_{u,d}$-$m_s$ plane \cite{Karsch:2001vs}.

Symmetry considerations help us to identify the order of the transition on
the corners of the diagram \cite{Pisarski:ms}.  
In the upper right hand corner we have pure
gauge theory with a deconfinement transition at $T\approx 270$ MeV.  In the
lower-left corner of the diagram, we expect a first order transition 
for three massless degenerate flavours.  Chiral symmetry is broken at low
temperature and restored at high temperature.
We expect that as the quark masses are increased, the
transition will weaken as we approach a second order line.  When the strange
quark is very heavy, but $u$ and $d$ are light, we expect a second order
transition for massless quarks.  Karsch's estimates of the chiral transition
temperatures in the corners are shown in the diagram.


Some recent
results from the MILC collaboration \cite{Bernard:2002yd}
with the Aqstad action and either
2+1 or 3 degenerate flavours indicate that above 175--200 MeV 
$\langle\bar\psi\psi\rangle$ extrapolates to 0 as $m_q \rightarrow 0$.
This is evidence that chiral symmetry is restored at higher temperature.
For 2+1 flavours, with $N_t=6$, the chiral symmetry restoration starts
at about 200 MeV.  The runs with $N_t=8$ continue as the lightest
quark mass runs are not completed.  Here it appears the transition may occur
at about 175 MeV.  The difference can be a finite lattice spacing effect.
Calculations have also been done with three degenerate quarks.  The lightest
quark mass explored by MILC is 0.4 $m_s$ where $m_s$ is the strange quark mass.
Each of the individual curves looks quite smooth, so it is not clear that
any of these masses are within the first order transition region;
however, if the curves are extrapolated to zero quark mass, there
appears to be chiral symmetry restoration around 185 MeV.

\begin{figure}
\begin{center}
\begin{tabular}{c c}
\includegraphics[width=2.35in]{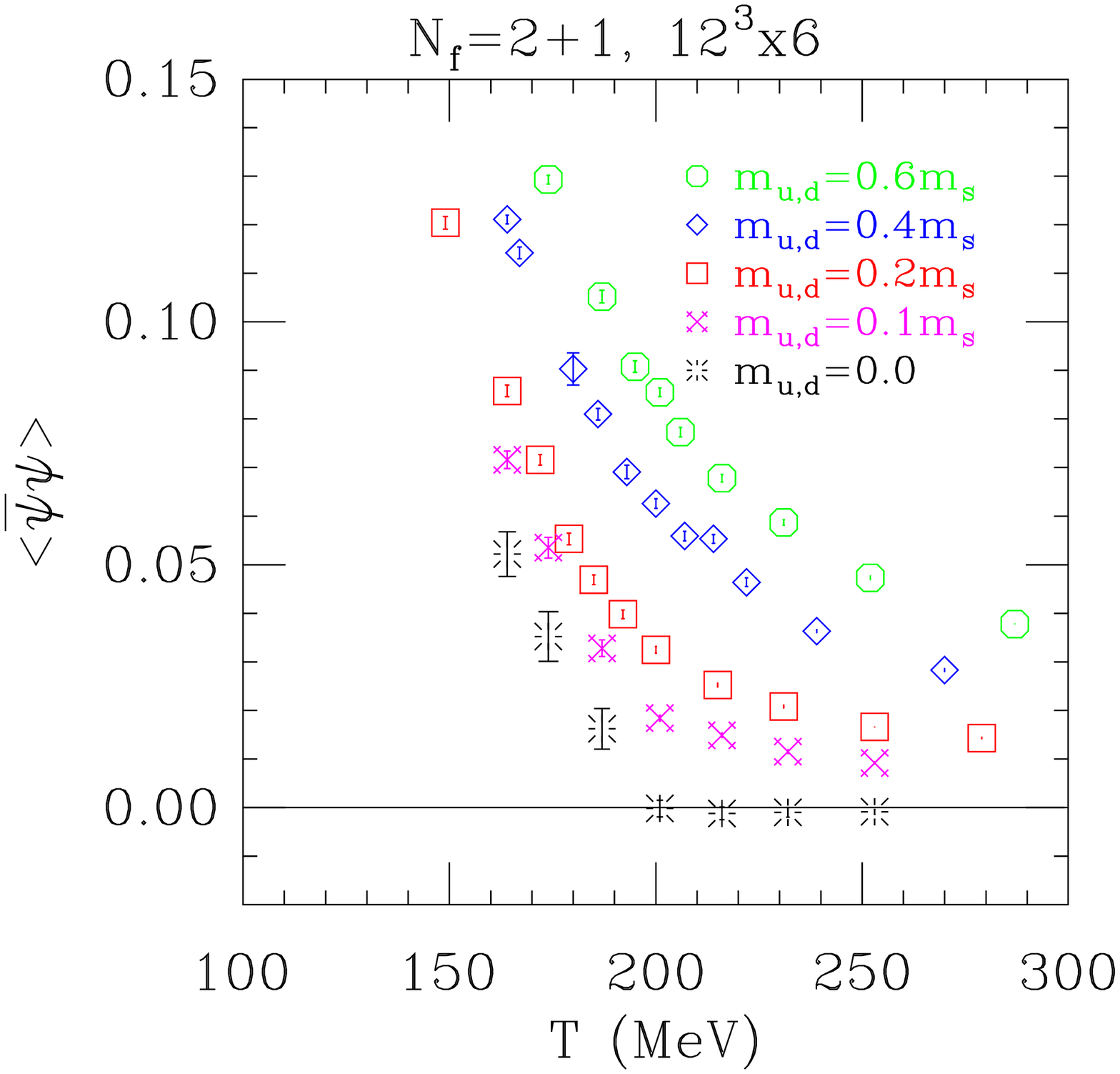}
&
\includegraphics[width=2.35in]{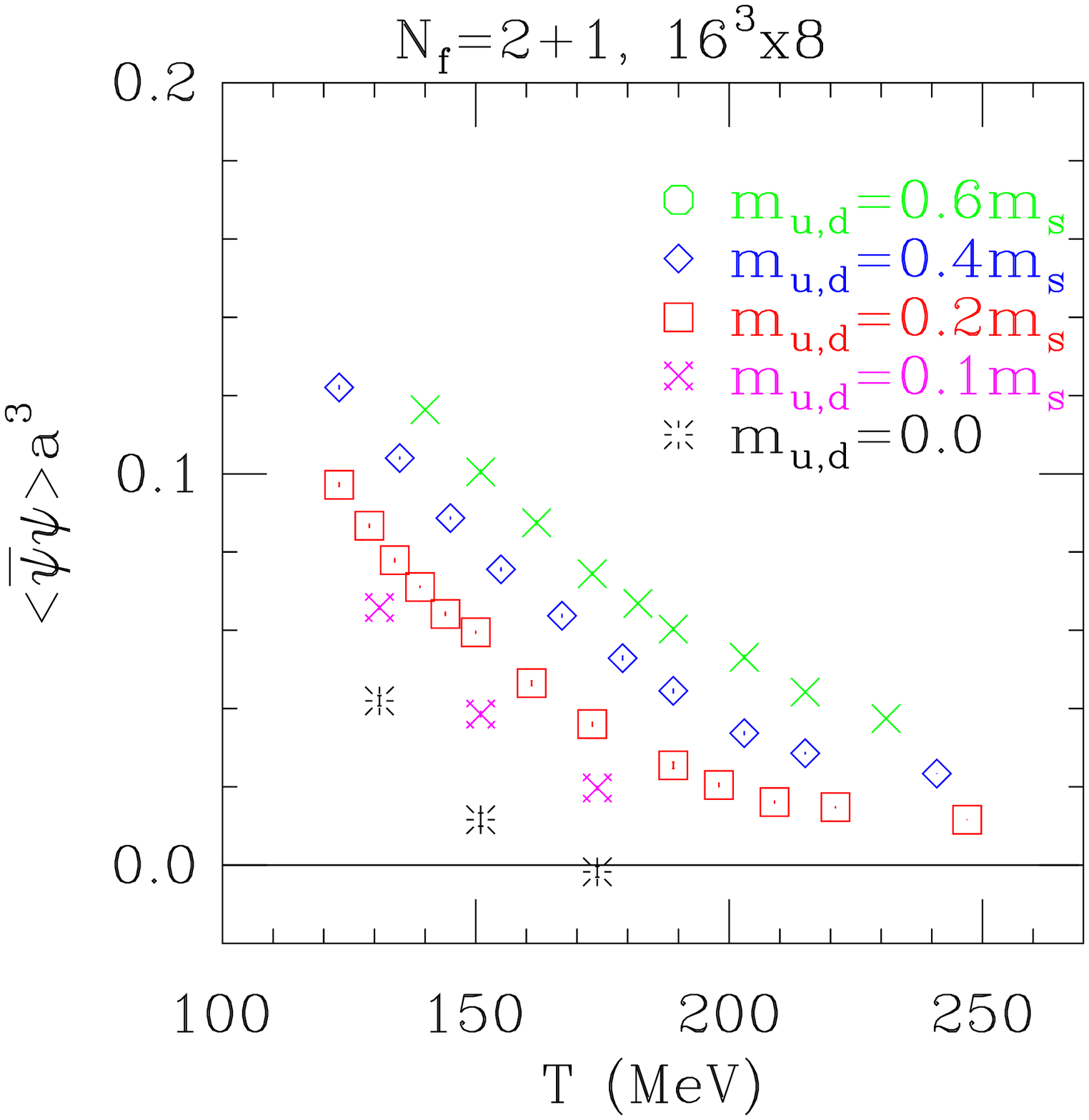}
\end{tabular}
\end{center}
\caption{$\langle\bar\psi\psi\rangle$ {\it vs}. temperature for various
light quark masses.  Bursts are extrapolations to zero quark mass.
These graphs are updated from the ones in reference~\cite{Bernard:2002yd}.
\label{threeflavor_milc}
}
\end{figure}

\begin{figure}[tb]
\epsfxsize=2.25in
\epsfysize=2.25in
\begin{center}
\epsfbox{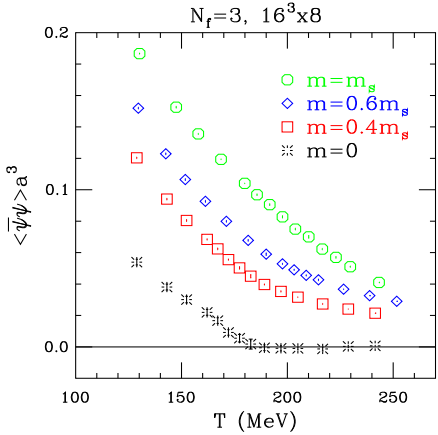}
\end{center}
\caption{$\langle\bar\psi\psi\rangle$ for three degenerate quarks with
$N_t=8$.  
}
\end{figure}

\begin{figure}[tb]
\epsfxsize=2.25in
\epsfysize=2.25in
\begin{center}
\epsfbox{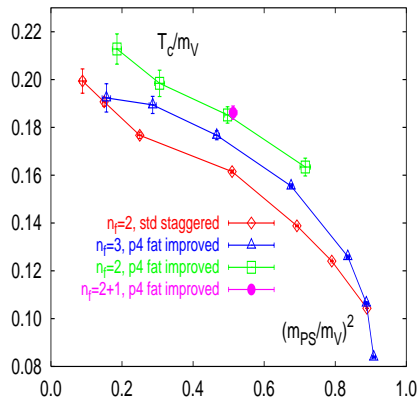}
\end{center}
\caption{Ratio of transition temperature to vector ($rho$) mass {\it vs}.
squared ratio of pseudoscalar to vector masses \cite{Karsch:2000kv}.  
Results are shown for
various numbers of dynamical quarks.  For $N_f=2$ results with an unimproved
action are shown for comparison.  
}
\label{tc_vs_ratio}
\end{figure}

Turning now to the P4 action, Karsch, Laermann and Peikert 
have done an extensive study with $N_t=4$ \cite{Karsch:2000kv}.
A graph summarizing their results for the ratio of $T_c$ to the vector 
meson mass is show in \fref{tc_vs_ratio}.  
For $N_f=2$ they show both the P4 action and
the standard KS action results.  Clearly, the improved action gives a higher
value for $T_c$.  With three degenerate flavours, the transition temperature
is lower than for $N_f=2$ (considering only improved action results).
One combination of quark masses is available for $N_f=2+1$; however, the
result looks very much like the $N_f=2$ result.  This could be because
the strange quark mass is so much larger than its physical value.
The Bielefelders estimate that in the chiral
limit $T_c= 173 \pm 8 $ MeV for $N_f=2$ and $154\pm8$ MeV for $N_f=3$.

We see that the two groups do not seem to be in close agreement 
on the transition temperature for $N_f=3$.
A careful comparison would be useful.  It is possible that the difference
is a finite lattice spacing effect.

We would like to find the line in the phase diagram separating the first
order transition region for very light quarks from the crossover region
for heavier quarks.

\begin{figure}[tb]
\epsfxsize=2.25in
\epsfysize=2.25in
\begin{center}
\epsfbox{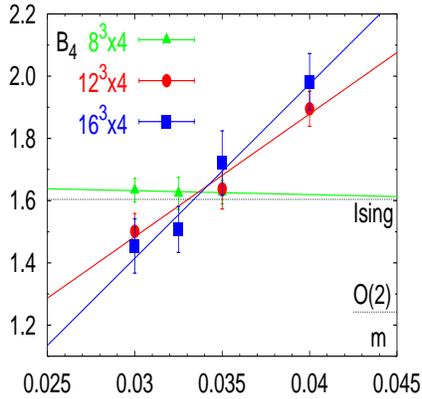}
\end{center}
\caption{Binder cumulant from Karsch, Laermann \& Schmidt 
\cite{Karsch:2001nf}.  Results from different
volumes should cross at the critical point separating the first order region
from the smooth crossover.
\label{binder}
}
\end{figure}

MILC is exploring a horizontal line at fixed $m_s$ and the diagonal with
$m_s=m_{ud}$.  
The results shown for $\langle\bar\psi\psi\rangle$ are rather smooth curves
except perhaps for the lightest quark mass studied on the $16^3\times8$
lattice.  MILC does not claim to have worked at light enough quark mass
to be in the first order transition region.  Two other groups have also
studied this question.
Karsch, Laermann \& Schmidt \cite{Karsch:2001nf} 
and Christ \& Liao \cite{Christ:lat02} 
have used light quarks with $N_t=4$ and the unimproved action.
Both groups are now using the Binder cumulant method to determine the order
of the transition.  In \fref{binder}, the Binder cumulant is shown as a function
of quark mass for three different spatial volumes.  The lines should
cross at the critical point, which appears to be about 0.035 in lattice units.

The two groups estimate the pseudoscalar mass at the critical point to be
270 and 290 MeV.  Karsch {\it et al}.\ have some results with the P4 action
for which $M_{ps} \approx 192(25)$ MeV.  Clearly, given the difference between
the unimproved and P4 actions, more work to control the
cutoff effects is needed to capitalize on these exploratory works.
Also, as interesting as is the three degenerate flavour case, we are most
interested in $m_s > m_{ud}$.

\begin{figure}
\begin{center}
\begin{tabular}{c c}
\includegraphics[width=2.25in]{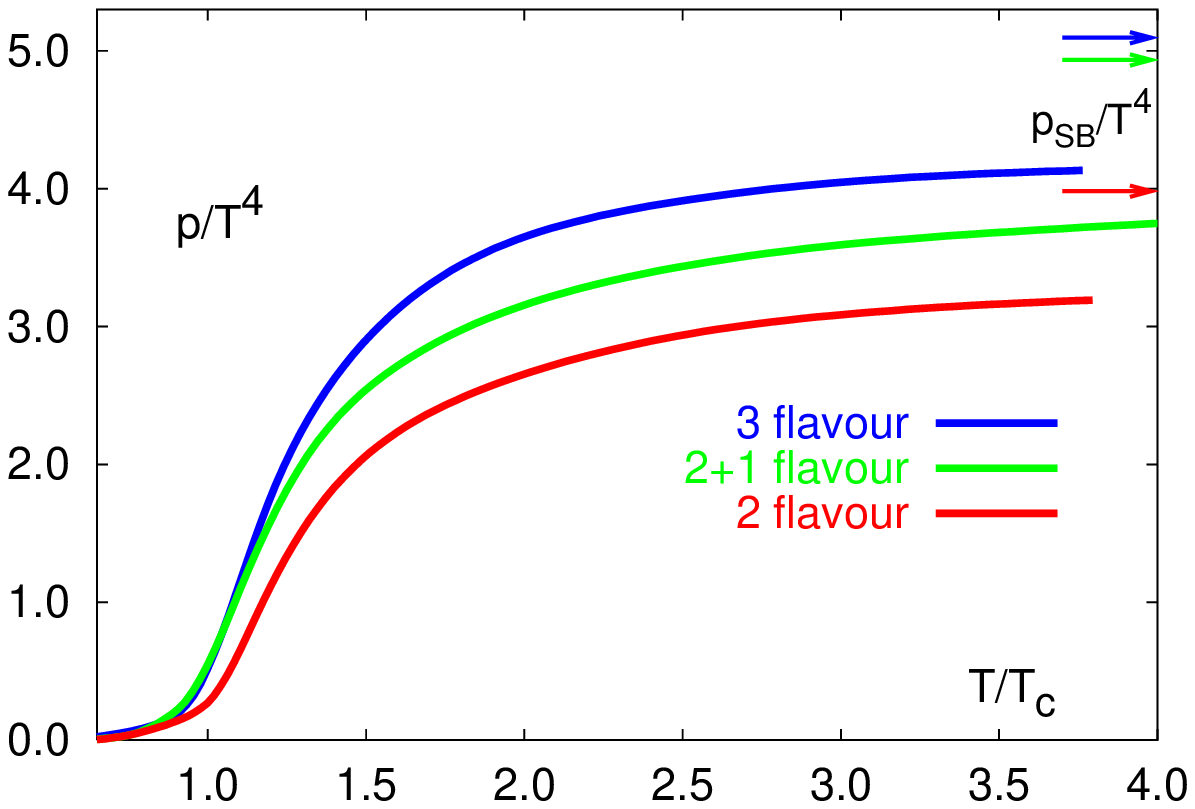}
&
\includegraphics[width=2.25in]{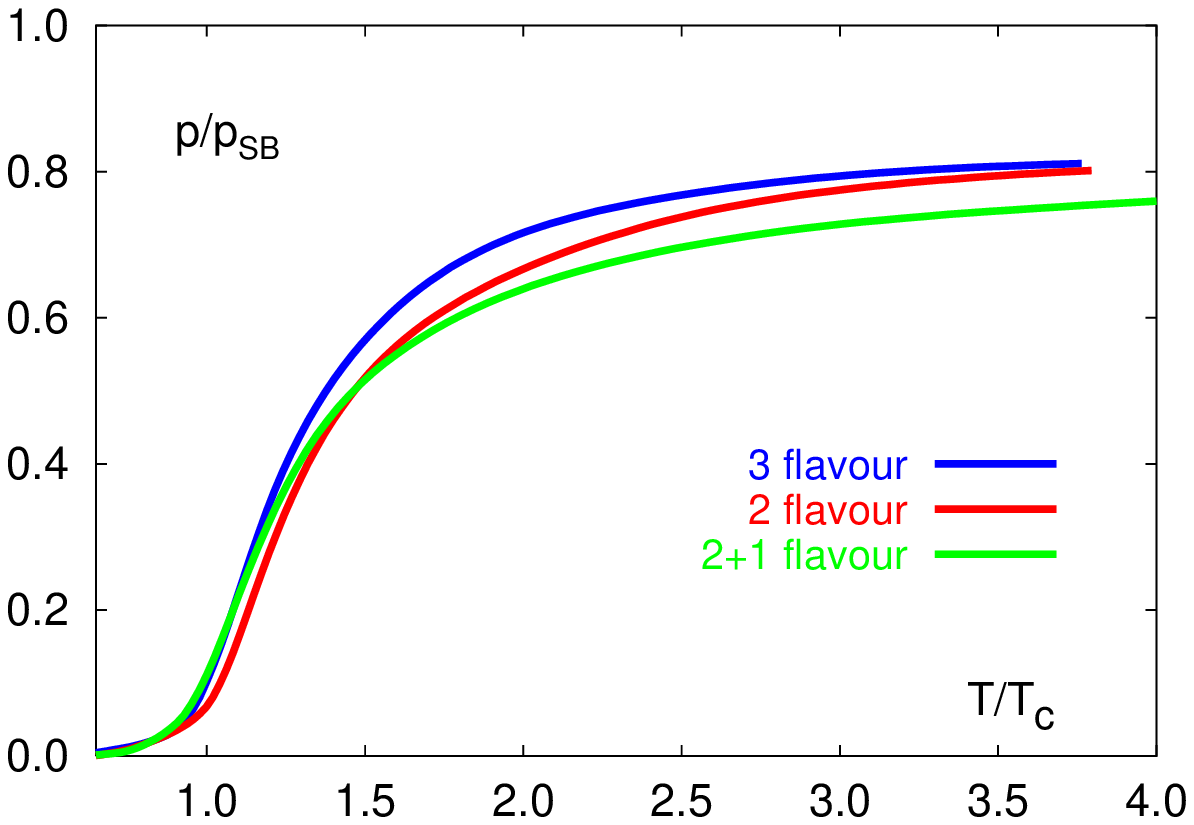}
\end{tabular}
\end{center}
\caption{Pressure for various numbers of flavours \cite{Karsch:2000ps}.  For $N_s=2$ and 3, 
$m_q=0.4T$. For 2+1, $m_{u,d}=0.4T$ and $m_s=T$.
\label{karschpressure}
}
\end{figure}
\begin{figure}[tb]
\epsfxsize=2.0in
\epsfysize=2.0in
\begin{center}
\epsfbox{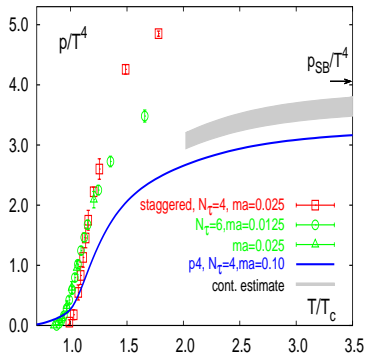}
\end{center}
\caption{Pressure estimate by Karsch, Laermann and Peikert \cite{Karsch:2000ps}
of continuum limit for
$T > 2 T_c$.
}
\label{continuumpressure}
\end{figure}

The pressure of the quark-gluon plasma
is of great interest.  Using the P4 action, Karsch,
Laermann and Peikert \cite{Karsch:2000ps} 
have done an extensive study for
$N_t=4$.  Good control of flavour symmetry is important near $T_c$ because 
there is a system of light hadrons, and flavour symmetry breaking distorts
the spectrum of the light hadrons.  Some fattening is used in this calculation
to improve flavour symmetry, but it would be necessary to repeat the
calculation with large $N_t$, i.e., smaller lattice spacing to get
reliable results near $T_c$.  However,
far above $T_c$, these results may be indicative of the continuum limit.
\Fref{karschpressure} shows the pressure using the P4 action for various 
combinations
of quark mass and flavours.  Indicated by arrows are the free quark values
at very high temperature.  In the second graph, each pressure curve 
is normalized by the corresponding value.  A comparison with unimproved
action results and an estimate of the continuum limit at high temperature
is in \fref{continuumpressure}.

\section{Strange Quark Content of QGP}

MILC has looked at various quark number susceptibilites \cite{Bernard:2002yd}.  
They have been related to event-by-event fluctuations in heavy ion collisions
in work by B.~Muller
\cite{Muller:2001wj} 
using the fluctuation-dissipation theorem.

\begin{equation}
\left\langle \delta Q^2 \right\rangle \propto \frac{T}{V_s} \frac{\partial^2
 \log Z}{\partial \mu^2_Q} = \chi_Q(T, \mu_Q=0) ~,
\end{equation}

We define:
\begin{equation}
\chi_{ij} = \frac{T}{V_s} \frac{\partial^2 \log Z}{\partial \mu_i
 \partial \mu_j} \Bigg|_{\mu=0} ~,
\end{equation}
\begin{equation}
\chi_{sing} = 2 \chi_{uu} + 2 \chi_{ud}
\end{equation}
and triplet quark number susceptibility
\begin{equation}
\chi_{trip} = 2 \chi_{uu} - 2 \chi_{ud} ~.
\end{equation}

\begin{figure}
\begin{center}
\begin{tabular}{c c}
\includegraphics[width=2.25in]{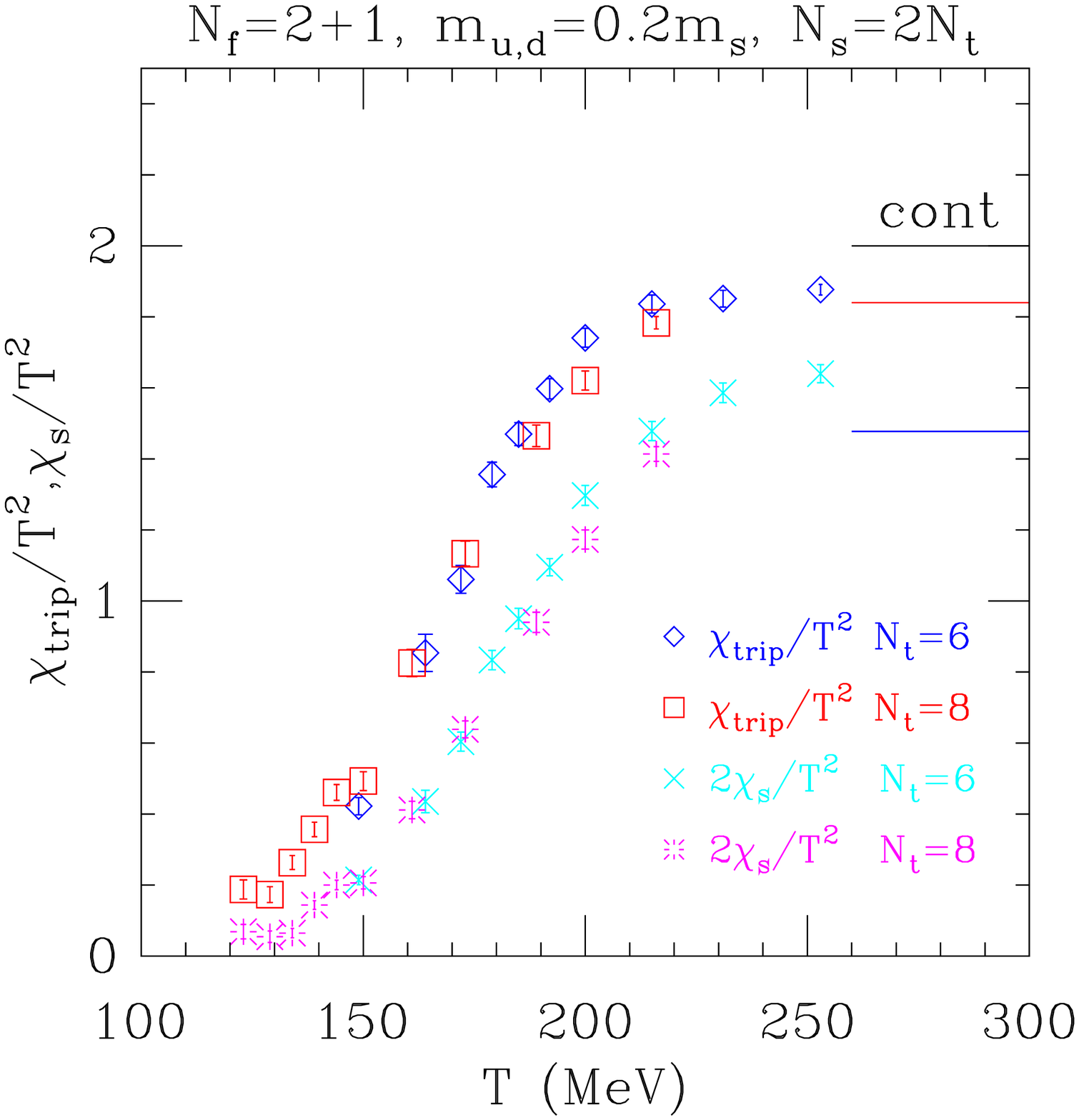}
&
\includegraphics[width=2.25in]{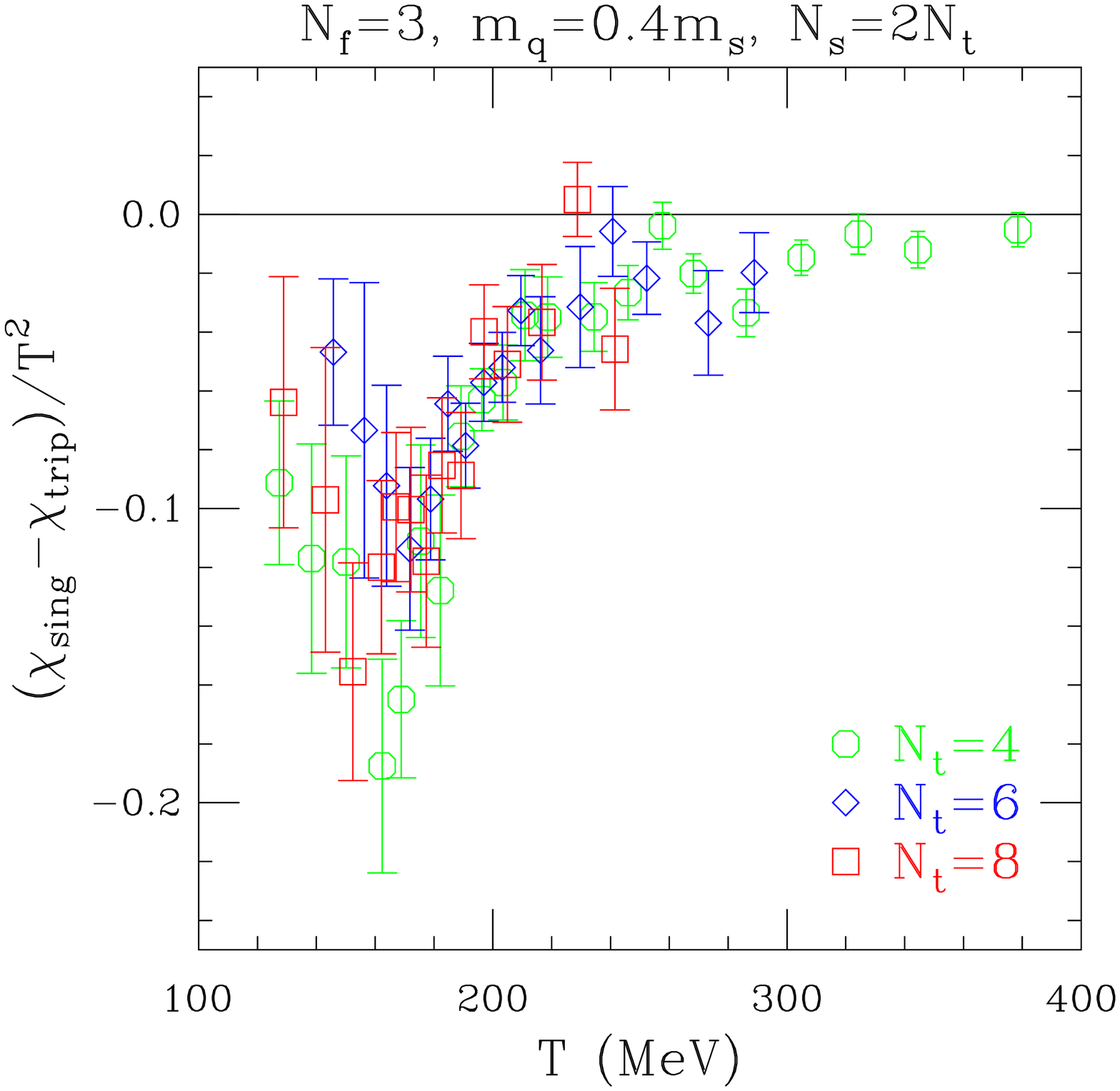}
\end{tabular}
\end{center}
\caption{Left: The triplet and strange quark number susceptibilities for
$N_f=2+1$ with $m_{u,d}=0.2\, m_s$ on $12^3\times 6$
and $16^3\times 8$ lattices \cite{Bernard:2002yd}. 
Right: The difference between singlet and triplet quark number
susceptibility for $N_f=3$ with quarks of mass $m_q=0.4\, m_s$, on
$8^3\times 4$, $12^3\times 6$ and $16^3\times 8$ lattices.
}
\label{milc_susceptibility}
\end{figure}

As seen in the second part of \fref{milc_susceptibility}, 
the difference between $\chi_{sing}$
and $\chi_{trip}$ is most pronounced around the transition temperature.
MILC also has results for the strange quark susceptibility.  It would be
useful to try to turn these results into a statement about the strangeness
content of the QGP in a way that is testable at RHIC.

\section{Nonzero Chemical Potential}

Due to lack of time, this material was not presented in my talk.  That
was unfortunate, because very significant progress has been made in
algorithms, and we can expect to see concomitant advances in calculations soon.

Calculations with finite chemical potential are important because experiments
at heavy-ion colliders such a RHIC
start with nuclei, not anti-nuclei so there is
a quark chemical potential $\mu$ of about 15 MeV \cite{Braun-Munzinger:2001ip}.
However, finite chemical potential is difficult because the action is no longer
real so it is no longer possible to implement importance sampling in
the usual way.  Recent progress has been made using three new methods.

\begin{figure}[tb]
\epsfxsize=2.50in
\epsfysize=2.50in
\begin{center}
\epsfbox{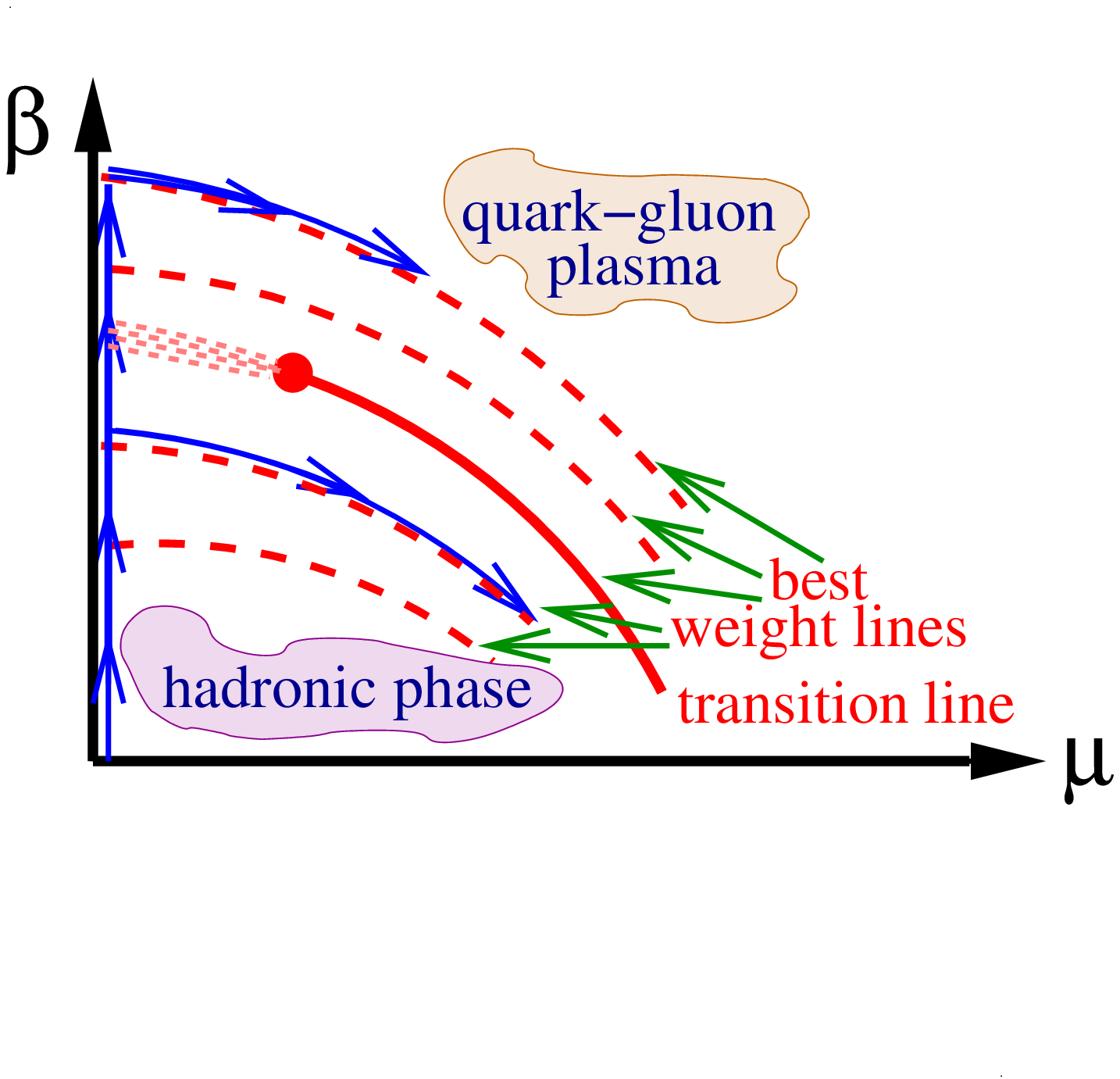}
\end{center}
\vspace{-0.5in}
\caption{
Best weight lines in $\mu-\beta$ plane \cite{Csikor:2002ic}.
}
\label{finite_mu_phase_diagram}
\end{figure}

Fodor and Katz introduced a multiparameter reweighting 
technique \cite{Fodor:2001au}.  
The key here is that the gauge coupling is shifted from $\beta$
to $\beta_0$ in order
to maximize the overlap of the ensemble generated with
$\mu=0$ with the desired $\mu\ne0$ ensemble.  This is detailed 
in \eref{reweight} and illustrated in \fref{finite_mu_phase_diagram}
\cite{Csikor:2002ic}.

\begin{eqnarray}
Z(\beta,\mu,m) = \int {\cal D}U\exp[-S_{g}(\beta)]\det M(\mu,m)
\nonumber \\
=\int {{\cal D}U \exp[-S_{g}(\beta_0)]\det M(\mu=0,m)} \label{reweight}
 \\
{ \times \left\{\exp[-S_{g}(\beta)+S_{g}(\beta_0)]
\frac{\det M(\mu,m)}{\det M(\mu=0,m)}\right\} },
\nonumber\end{eqnarray}

A second method introduced by a Swansea-Bielefeld group uses a Taylor series
expansion in $\mu/T$ of both the reweighting factor and observable variables
\cite{Allton:2002zi}.
Thus, reweighting is avoided but the chemical potential must be small as only
the leading order term is available.

de Forcrand and Philipsen use analytic continuation from imaginary $\mu$
\cite{deForcrand:2002ci,deForcrand:2002yi}.
When the chemical potential is imaginary, the action is real and importance
sampling works.  The challenge is to do the analytic continuation.  It will
be interesting to see which of these approaches is best.  We are certainly
learning more about QCD with $\mu\ne0$ than we had previously been able to.

\section{Prospects}

I hope I have convinced you that there are {\it many}
interesting calculations being done at finite temperature with $N_f=2+1$ and 3.
However, a great deal remains to be done:
we need better control of systematic errors, including the
continuum limit and wider coverage of the phase diagram.
I have concentrated on results with improved KS type quarks, but these
are not the only methods for quarks.  For example, 
improved actions for Clover quarks are being pursued by the CP-PACS/JLQCD
collaborations \cite{Aoki:2002vh}.  
Dynamical quark calculations with domain wall or overlap
quarks and calculations with chemical potential are just in their infancy.
Although we lattice practitioners
have learned a great deal recently, there is much we can
and {\it shall\/} do to improve our calculations.

\ack
I am very grateful to the organizers, particularly Steffen Bass and Berndt
Muller for inviting me to give this talk.  I would like to thank all my
MILC collaborators, especially Urs Heller and Bob Sugar
for reading this manuscript.  Juergen Engels, Frithjof Karsch, Edwin Laermann
and Bengt Petersson who have contributed so much to the study of QCD at
finite temperature, I thank for many stimulating conversations, hospitality
and friendship over the years.  Owe Philipsen and Peter Petreczky
provided excellent scientific companionship during the conference.

\end{document}